\documentclass[prd, preprint, nofootinbib]{revtex4}%

\usepackage{amsfonts}
\usepackage{amsmath}
\usepackage{amssymb}
\usepackage{mathrsfs}
\usepackage{amsfonts}

\usepackage{graphicx}
\usepackage{appendix}
\usepackage{color}

\newcommand{\dd}{{\rm{d}}} 
\newcommand{\rovno}{\!& = &\!}

\newcommand{\defeq}{:=}

\def\scri{{\!\mathscr{J}}}
\def\scrisub{{_{\!\mathscr{J}}}}

\newcommand{\im}{\mathrm{i}}

\begin{document}

\title{New metric useful for studies of gravitational radiation from accelerating black holes of type D with any cosmological constant \\[10mm]}

\author{Ji\v{r}\'i Podolsk\'{y}}
\email{jiri.podolsky@matfyz.cuni.cz}
\affiliation{Institute of Theoretical Physics,
Charles University, Faculty of Mathematics and Physics\\
V~Hole\v{s}ovi\v{c}k\'ach 2, 18000 Prague 8, Czechia}

\date{\today{}}



\begin{abstract}
\ \\[20mm]
We apply a convenient parametrization of all Pleba\'nski-Demia\'nski black holes of algebraic type~D with the Kerr rotation, NUT twist, acceleration, electric and magnetic charges, and any value of the cosmological constant~$\Lambda$, recently found by Astorino. This represents a complete class of exact solutions to the Einstein-Maxwell-$\Lambda$ equations such that the (non-null) electromagnetic field is aligned with both the (geodesic, shear-free, double) principal null directions of the Weyl tensor. We derive a novel appropriate metric form and the corresponding canonical frames naturally adapted to both the de~Sitter-like (${\Lambda>0}$) and the anti-de~Sitter-like (${\Lambda<0}$) conformal infinity, and we demonstrate that these geometric objects are useful for investigation of radiation properties. In particular, using the Fern\'andez-\'Alvarez-Senovilla criterion, it confirms that the gravitational radiation is present if, and only if, these black holes are accelerating.
\end{abstract}
\maketitle


\newpage

\section{Introduction}

The main purpose of this paper is to derive an explicit simple metric that is convenient for investigation of the asymptotic structure of the \emph{whole} family of Pleba\'nski-Demia\'nski (PD) black holes with \emph{any} value of the cosmological constant~$\Lambda$. Then we demonstrate that it can be used to test the criterion of the presence of gravitational radiation which was introduced in the context of recent works \cite{FernandezSenovilla:2020a, FernandezSenovilla:2020b, FernandezSenovilla:2022a, FernandezSenovilla:2022b, FernandezSenovilla:2026} by Fern\'andez-\'Alvarez and Senovilla.

The large Pleba\'nski-Demia\'nski class of exact solutions to the Einstein-Maxwell-$\Lambda$ field equations involves many famous black holes, namely the spherically symmetric Schwarzschild, Reissner-Nordstr\"{o}m (with charge), or  Kottler-Weyl-Trefftz (with~$\Lambda$) solutions. Moreover, it contains their  axisymmetric rotating extensions, in particular those of Kerr, Newman-Unti-Tamburino, and Kerr-Newman. Such spacetimes are static or stationary, so it is expected that they do not contain gravitational and electromagnetic radiation. Interestingly, the Pleba\'nski-Demia\'nski class also contains their \emph{accelerating generalizations}. The first of them, the so-called C-metric, was discovered by Levi-Civita and Weyl, and then studied by Ehlers and Kundt, Kinnersley and Walker, Bonnor, and many others (see \cite{Stephanietal:2003, GriffithsPodolsky:2009} for a review). The full PD family of spacetimes may thus serve as an important test of the Fern\'andez-\'Alvarez-Senovilla approach, because such black holes combine possible acceleration with any value of the cosmological constant (and other physical parameters).

The PD class is geometrically defined as spacetimes of algebraic type~D  whose both principal null directions (PNDs) are geodesic and shear-free. They admit any cosmological constant, and electromagnetic field that is non-null and double-aligned with the PNDs of the Weyl tensor. Generalizing the non-accelerating solutions of this type given by Carter \cite{Carter:1968b}, and all type~D vacuum spacetimes identified by Kinnersley \cite{Kinnersely:1969a}, such a complete family was found in 1971 by Debever \cite{Debever:1971}, but its more convenient metric was presented by Pleba\'nski and Demia\'nski \cite{PlebanskiDemianski:1976}  in 1976. In a series of works \cite{GriffithsPodolsky:2005, GriffithsPodolsky:2006, PodolskyGriffiths:2006} in 2006 the whole PD class was rewritten by Griffiths and Podolsk\'y (GP) into the form which is more suitable for direct identification of all its subclasses (with no acceleration, no twist, etc.), enabling various mathematical and physical studies of these exact black holes  (see \cite{Stephanietal:2003, GriffithsPodolsky:2009, VandenBergh:2017} for more details). This GP metric was further improved a few years ago by Podolsk\'y and Vr\'atn\'y (PV) in~\cite{PodolskyVratny:2021, PodolskyVratny:2023}.

However, even these better representations of the complete Pleba\'nski-Demia\'nski family contained a subtle peculiarity: They did not identify accelerating black holes with just the NUT parameter and vanishing Kerr-like rotation, so it was  misleadingly claimed that these must be of a general algebraic type. It thus came as a surprise when Astorino~(A) in 2024 presented a novel metric form of type~D spacetimes, obtained by specific solution generating techniques \cite{Astorino:2024b, Astorino:2024a}. Not only it explicitly includes the ``missing'' accelerating NUT black holes, but it admits \emph{direct} limits to \emph{all} the subcases. This is achieved simply by setting (any of) its physical parameters to zero, namely $m, a, l, \alpha, e, g, \Lambda$, which represent mass, Kerr rotation, NUT twist parameter, acceleration, electric charge, magnetic charge, cosmological constant, respectively.

In recent works \cite{OvcharenkoPodolskyAstorino:2025a, OvcharenkoPodolskyAstorino:2025b} we proved that the new Astorino solution is fully equivalent to the Pleba\'nski-Demia\'nski solution. To achieve it, we first put the original A metric into a more compact A$^+$ form, and then elucidated its relations to PD, GP, and PV metric representations, identifying all the different coordinates and parameters.

The Pleba\'nski-Demia\'nski class of exact spacetimes is a nice playground for investigation of a nontrivial interplay between the (possibly rotating, charged, and NUTed) \emph{black holes} and their \emph{gravitational radiation} in the \emph{cosmological} setting with nonzero~$\Lambda$. For ${\Lambda=0}$ these spacetimes contain asymptotically flat regions, and standard methods developed since 1950s can thus be applied --- they confirmed that accelerating black holes (represented by the C-metric) indeed emit specific gravitational radiation. However, in the ${\Lambda\ne0}$ case the situation is much less satisfactory due to the lack of a generally accepted rigorous definition of gravitational waves in this more general cosmological context. The recent approach by Fern\'andez-\'Alvarez and Senovilla, introduced an elaborated in their works \cite{FernandezSenovilla:2020a, FernandezSenovilla:2020b, FernandezSenovilla:2022a, FernandezSenovilla:2022b, FernandezSenovilla:2026}, is thus welcome and inspiring. It arises from the analysis of the (conformally rescaled version of the) Bel-Robinson tensor, evaluated near the conformal infinity which can be de~Sitter-like and anti-de~Sitter-like. It determines the asymptotic supermomentum, that naturally decomposes into the super-Poynting vector and the superenergy density.  Because the Bel-Robinson tensor is a \emph{quadratic combination} of the Weyl tensor components (the asymptotic super-Poynting vector is a \emph{commutator} of its ``electric'' and ``magnetic'' parts), it overcomes some previous limitations of studying the behavior of asymptotically de Sitter and anti-de~Sitter spaces using only the Weyl tensor itself, such as in \cite{KrtousPodolsky:2003, KrtousPodolskyBicak:2003, PodolskyOrtaggioKrtous:2003, KrtousPodolsky:2004, KrtousPodolsky:2005,PodolskyKadlecova:2009}.

We already applied this general, covariant and gauge-invariant framework to the PD class of black holes in the work \cite{FernandezPodolskySenovilla:2024}, employing the PV representation introduced in~\cite{PodolskyVratny:2023}. The Fern\'andez-\'Alvarez-Senovilla criterion of the presence of gravitational radiation was successfully tested, and confirmed. However, this remained restricted to the ${\Lambda>0}$ spacetimes only, and the twist parameters $a$ and $l$ were not properly identified for the most general PD black holes having all 7 parameters.

This paper aims to remedy these two constraints. We start with the \emph{most convenient A$^+$ representation} of all Pleba\'nski-Demia\'nski black holes, and we admit \emph{any value of the cosmological constant~$\Lambda$}. The results are thus fully general, with both ${\Lambda>0}$ and ${\Lambda<0}$, and the physical parameters of the black holes can be (independently, and in any order) set to zero. In particular, it turns out that the acceleration parameter~$\alpha$ uniquely determines the presence or absence of gravitational radiation in this large class of exact black holes.

The work is organized as follows. In Sec.~\ref{sectionc:new-metric} we derive the new metric suitable for the analysis of global structure of the whole class of type~D black holes. In Sec.~\ref{sections:scri-and-hab} we localize the conformal infinity, calculate its normal, and derive the conformal metric on $\scri$. The fundamental frames on $\scri$, adapted to PNDs, are explicitly introduced in Sec.~\ref{section:PND-Weyl-for-Lambda>0} and Sec.~\ref{section:PND-Weyl-for-Lambda<0} for ${\Lambda>0}$ and ${\Lambda<0}$, respectively. Their application to the study of gravitational radiation for ${\Lambda>0}$ and ${\Lambda<0}$, by evaluating the corresponding super-Poynting vector and the superenergy density, is contained in final Sec.~\ref{section:sP-for-Lambda>0} and Sec.~\ref{section:sP-for-Lambda<0}.

\newpage

\section{New convenient form of the metric}
\label{sectionc:new-metric}

In \cite{Astorino:2024b}, Marco Astorino presented a novel metric representation of the most general type~D black hole with any cosmological constant $\Lambda$ and a non-null electromagnetic field that is
doubly-aligned with the (geodesic and shear-free) principal null directions of the Weyl tensor. Subsequently, in \cite{OvcharenkoPodolskyAstorino:2025a,OvcharenkoPodolskyAstorino:2025b} we put this Astorino metric into a more compact form, denoted as~A$^+$, namely
\begin{equation}
    \dd \hat{s}^2=\dfrac{1}{\Omega^2}\bigg[-\dfrac{\Delta_r}{\rho^2}(A\,\dd t' - B\,\dd\varphi')^2
    + \dfrac{\Delta_x}{\rho^2}(C\,\dd t' + D\,\dd\varphi')^2
    + \frac{\rho^2}{1+\alpha^2a^2} \Big(\,\dfrac{\dd r^2}{\Delta_r} + \dfrac{\dd x^2}{\Delta_x}\,\Big)\bigg],
    \label{ds2_simpl}
\end{equation}
where the functions are\footnote{Following \cite{Astorino:2024b,OvcharenkoPodolskyAstorino:2025b},
 we set the additional dimensionless constant $C_f$ to the specific value ${1/(1+\alpha^2a^2)}$.}
\begin{align}
    \Omega(r,x) &= 1-\alpha\, r\, x \,,\label{Om_cf}\\
    A(x) &= 1 + \alpha^2(l^2-a^2)\,x^2\,,\label{Aw}\\
    B(x) &= a + 2l\,x + a\,x^2\,,\label{Bw}\\
    C(r) &= a + 2\alpha l\,r + \alpha^2 a\, r^2\,,\label{Cw}\\
    D(r) &= (l^2-a^2) + r^2\,,\label{Bu}
\end{align}
and the slightly more involved polynomials read
\begin{align}
\rho^2(r,x) =&\  AD+BC \label{rho2expl}\\
       =&\  (l+a\,x)^2 + 2\alpha\,l\,(a+2l\,x+a\,x^2)\,r + \alpha^2(a^2-l^2)^2\,x^2 + [1+\alpha^2(a+l\,x)^2\,]\,r^2 , \nonumber\\[2mm]
\Delta_r(r)=&\ (1-\alpha^2r^2)\big[(r-m)^2 - (m^2+l^2-a^2-e^2-g^2)\big] \nonumber\\
   & \hspace{20mm} - \frac{\Lambda}{3}\Big(\,\frac{3l^2}{1+\alpha^2a^2}\,r^2
    + \frac{4\alpha a l}{1+\alpha^2a^2}\,r^3 + r^4 \Big) ,\label{delta_r_init} \\[2mm]
\Delta_x(x)=&\ (1-x^2)\big[(1-\alpha m \,x)^2-\alpha^2x^2(m^2+l^2-a^2-e^2-g^2)\big] \nonumber\\
   & \hspace{20mm} - \frac{\Lambda}{3}\Big(\,\frac{3l^2}{1+\alpha^2a^2}\,x^2
    +\frac{4al}{1+\alpha^2a^2}\,x^3 + \frac{a^2+\alpha^2(a^2-l^2)^2}{1+\alpha^2a^2}\,x^4 \Big) . \label{delta_x_init}
\end{align}
Here $m$~is the \emph{mass} parameter, $a$~is the Kerr-like \emph{rotation}, $l$~is the \emph{NUT parameter}, $\alpha$~is \emph{acceleration}, $e$~and $g$~are \emph{electric} and \emph{magnetic charges}, respectively, and $\Lambda$~denotes the \emph{cosmological constant}. As usual, the coordinates $t', r$ have the physical dimension of length, while $x, \varphi'$ are dimensionless. The quantities  $m, a, l, e, g$ also have the physical dimension of length, while the dimension of~$\alpha$ and $\sqrt{\Lambda}$~is the inverse length.

This explicit form of the metric has a very nice property, namely that \emph{all subclasses} of the complete family of Pleba\'nski-Demia\'nski black holes can be \emph{directly obtained by simply setting the corresponding parameter to zero}. For example, accelerating Kerr-Newman-(A)dS black holes are obtained for ${l=0}$, they further reduce to non-accelerating black holes when ${\alpha=0}$, etc. Keeping just $m$, the spherically symmetric Schwarzschild black hole is recovered because  ${\Omega = 1}$, ${A = 1}$, ${D = r^2}$, ${B = 0 = C}$, ${\rho^2 = r^2}$, ${\Delta_r = r^2-2mr}$, ${\Delta_x = 1-x^2}$, so it suffices to introduce ${x=\cos\theta}$.

Recall also that for ${e,g\ne0}$ the black holes are charged, with the aligned electromagnetic field given by the vector potential ${A(r,x)=A_{t'}\,\dd t' + A_{\varphi'}\,\dd\varphi'}$, where
\begin{align}
 A_{t'} =& \,\sqrt{\frac{1+\alpha^2(a^2-l^2)}{1+\alpha^2a^2}}\,\frac{-1}{\rho^2\, l}
       \Big[ \big[\,g (r + \alpha a l +  \alpha^2 a^2 r) + e l \,\big]\,r  \nonumber\\
 & \hspace{3mm} + l\,(g-\alpha a e) (a+2 \alpha l\, r+\alpha^2a\,r^2)\,x
  \label{At}\\
 & \hspace{3mm} +(a + \alpha l\,r)\big[\,ag + \alpha l(gr-el) + \alpha^2a\,[g(a^2-l^2)-el\,r] \big]\,x^2\Big]
   + \frac{g}{l} \,,  \nonumber \\
 A_{\varphi'} =& \,\sqrt{\frac{1+\alpha^2(a^2-l^2)}{1+\alpha^2a^2}}\,\frac{x}{\rho^2}
         \Big[
    \alpha l^3 (e+\alpha ag)x  + l^2\big[\alpha^2a(e+\alpha a g)\,xr + (\alpha a e-g)\big] \label{Aphi}\\
 & \hspace{3mm}+ (1+\alpha^2a^2)\big[ ae(x+\alpha\,r)\,r - gal\,x + g(r-\alpha a^2x)\,r + el(2+\alpha xr)r \big]  \Big] - a\, A_{t'}\,,  \nonumber
\end{align}
see \cite{OvcharenkoPodolskyAstorino:2025b}.

Since for ${\alpha \ne 0}$ the black holes accelerate, it is expected that they generate specific electromagnetic and gravitational radiation. However, rigorous investigation of such a radiation is not simple because the conformal infinity, located at ${\Omega=0}$, is determined by the condition ${r = 1 / (\alpha\, x)}$. Not only this depends on the spatial coordinate $x$ (that is on $\theta$), but it requires a delicate limit ${\alpha \to 0}$.

To resolve this problem, here we introduce a \emph{new metric form} associated with \eqref{ds2_simpl}, in which we first relabel the metric functions ${\Omega, \rho, C, D}$ to ${\Omega', \rho', C', D'}$, and then perform the coordinate transformations
\begin{equation}
  q = \frac{1}{r}\,,
  \label{q=1/r}
\end{equation}
and
\begin{equation}
  t = \frac{t'}{\sqrt{1+\alpha^2a^2}}\,,\qquad
  \varphi = \frac{\varphi'}{\sqrt{1+\alpha^2a^2}}\,.\label{t-phi-scaling}
\end{equation}
By introducing the metric functions by the specific rescaling
\begin{equation}\label{rescaled-funtions}
\Omega \defeq    q\, \Omega' \,, \quad
\rho^2 \defeq    q^2 \rho'^2 \,, \quad
C   \defeq    q^2 C' \,, \quad
D   \defeq    q^2 D' \,,
\end{equation}
\begin{align}
{P} \defeq (1+\alpha^2a^2)\, \Delta_x  \,,\qquad
{Q} \defeq (1+\alpha^2a^2)\,q^4 \Delta_r  \,,
\label{tilde-QP}
\end{align}
we obtain the metric
\begin{equation}
    \dd \hat{s}^2=\dfrac{1}{\Omega^2}\bigg[-\dfrac{{Q}}{\rho^2}(A\,\dd t - B\,\dd\varphi)^2
    + \dfrac{{P}}{\rho^2}(C\,\dd t + D\,\dd\varphi)^2
    + \rho^2\Big(\,\dfrac{\dd q^2}{{Q}} + \dfrac{\dd x^2}{{P}}\,\Big)\bigg],
    \label{A-new-metric}
\end{equation}
with the simple functions
\begin{align}
    \Omega &= q-\alpha\, x \,,\label{Omega_new}\\
    A &= 1 + \alpha^2(l^2-a^2)\,x^2\,,\label{A_new}\\
    B &= a + 2l\,x + a\,x^2,\label{AB_new}\\
    C &= \alpha^2 a + 2\alpha l\,q + a\, q^2\,,\label{C_new}\\
    D &= 1 + (l^2-a^2)\,q^2\,,\label{CB_new} \\
    \rho^2 &= AD+BC \,, \label{rho2_new}
\end{align}
and
\begin{align}
{P}(x) &= (1+\alpha^2a^2)(1-x^2)\big[\,1-2\alpha m \,x + \alpha^2(a^2+e^2+g^2-l^2)\,x^2\big] \nonumber\\
   & \hspace{10mm} - \frac{\Lambda}{3}\,
    \big(3l^2\,x^2 + 4al\,x^3 + [a^2+\alpha^2(a^2-l^2)^2]\,x^4 \big) , \label{P_new} \\[2mm]
{Q}(q) &= (1+\alpha^2a^2)(q^2-\alpha^2)\big[\,1-2m\,q + (a^2+e^2+g^2-l^2)\,q^2\big] \nonumber\\
   & \hspace{10mm} - \frac{\Lambda}{3}\,
   \big(1 + \alpha^2 a^2 + 4\alpha a l\,q + 3l^2\,q^2  \big) . \label{Q_new}
\end{align}
These are just polynomial expressions, both in the coordinates $x, q$ and in the physical parameters $m, a, l, \alpha, e, g$, and~$\Lambda$.

\section{Conformal boundary $\scri$ and its conformal metric $h_{ab}$}
\label{sections:scri-and-hab}

For conformally related metrics, ${g_{\alpha\beta} = \Omega^2\,\hat{g}_{\alpha\beta}}$, the \emph{conformal infinity}~$ \scri $ is located at
	\begin{equation}
	  \Omega = 0\,.
	  \label{Omega=0}
	\end{equation}
Here, the physical metric $\hat{g}_{\alpha\beta}$ is \eqref{A-new-metric}.	In view of \eqref{Omega_new}, i.e. ${\Omega = q-\alpha\, x}$, the conformal infinity is given by a very simple condition, namely
	\begin{equation}
	 \scri \hbox{ is located at:}\qquad    q = \alpha\,x\,.
	  \label{scri-AB}
	\end{equation}
For black holes without acceleration (${\alpha=0}$), the conformal infinity $ \scri $ is located at ${q = 0}$, that is at ${r=\infty}$. This is the standard result for the Kerr-Newman-(anti-)de Sitter  and also for NUT-(anti-)de Sitter black holes.\\

\subsection{The metric on $\scri$}
\label{subsectionc:hab}

Now we derive the \emph{metric $h_{ab}$ on the conformal boundary $ \scri $}. We start with the unphysical metric~$g_{\alpha\beta}$ corresponding to \eqref{A-new-metric}, namely the line element
\begin{equation}
    \dd s^2 = -\dfrac{{Q}}{\rho^2}(A\,\dd t - B\,\dd\varphi)^2
    + \dfrac{{P}}{\rho^2}(C\,\dd t + D\,\dd\varphi)^2
    + \rho^2\Big(\,\dfrac{\dd q^2}{{Q}} + \dfrac{\dd x^2}{{P}}\,\Big)\,.
    \label{conformal-metricJP2024}
\end{equation}
Evaluation on $ \scri $ of the metric functions \eqref{A_new}--\eqref{P_new}, using the condition \eqref{scri-AB}, yields explicit expressions (we use a subscript $\scri$ to denote the restriction of any object to $\scri$)
\begin{align}
    A_\scrisub &= 1 + \alpha^2(l^2-a^2)\,x^2,\qquad
    B_\scrisub  = a + 2l\,x + a\,x^2\,,\label{AB_on_scri}\\
    C_\scrisub &= \alpha^2 B_\scrisub \,,\hspace{28.3mm}
    D_\scrisub  = A_\scrisub \,, \label{CD_on_scri}\\
    \rho^2_\scrisub &= A_\scrisub^2 + \alpha^2 B_\scrisub^2 \,, \label{rho2_on_scri}
\end{align}
and
\begin{align}
{P}_\scrisub &= (1+\alpha^2a^2)(1-x^2)\big[\,1-2\alpha m \,x + \alpha^2(a^2+e^2+g^2-l^2)\,x^2\big] \nonumber\\
   & \hspace{10mm} - \frac{\Lambda}{3}\,
    \big(3l^2\,x^2 + 4al\,x^3 + [a^2+\alpha^2(a^2-l^2)^2]\,x^4 \big), \label{P_on_scri} \\[2mm]
{Q}_\scrisub &= \alpha^2 (1+\alpha^2a^2)(x^2-1)\big[\,1-2\alpha m\,x + \alpha^2(a^2+e^2+g^2-l^2)\,x^2\big] \nonumber\\
   & \hspace{10mm} - \frac{\Lambda}{3}\,
   \Big(1 + \alpha^2 a^2 + 4\alpha^2 a l\,x + 3\alpha^2l^2\,x^2 \Big) .\label{Q_on_scri}
\end{align}
By comparing \eqref{P_on_scri} with \eqref{Q_on_scri}, and using \eqref{AB_on_scri}--\eqref{rho2_on_scri}, we obtain an \emph{important identity} which will be crucial in our subsequent studies, namely
\begin{eqnarray}
 \big( {Q} + \alpha^2 {P} \big)_\scrisub = -\frac{\Lambda}{3}\,\rho^2_\scrisub  \,. \label{Q-on-scri-short}
\end{eqnarray}

In particular, this enables us to simplify the last term in the conformal (unphysical) metric (\ref{conformal-metricJP2024}) as
\begin{align}
\rho^2_\scrisub \Big(\,\dfrac{\dd q^2}{{Q}} + \dfrac{\dd x^2}{{P}}\,\Big)_{\!\scrisub} =
   -\frac{\Lambda}{3}\,\Big( \frac{\rho^4}{{P}{Q}} \Big)_{\!\scrisub}\,\dd x^2 \,. \label{conformal-metric-simplification-1}
\end{align}
Moreover, using also \eqref{AB_on_scri}--\eqref{rho2_on_scri}, the first two terms of \eqref{conformal-metricJP2024} combine on $\scri$ to
\begin{eqnarray}
&-\dfrac{{Q}_\scrisub}{\rho^2_\scrisub}(A_\scrisub\dd t - B_\scrisub\dd\varphi)^2 + \dfrac{{P}_\scrisub}{\rho^2_\scrisub}(C_\scrisub\dd t + D_\scrisub\dd\varphi)^2 \nonumber\\[2mm]
 & \hspace{25mm} = \dfrac{\Lambda}{3}\,(A_\scrisub\dd t - B_\scrisub\dd\varphi)^2
   + {P}_\scrisub(\alpha^2 \dd t^2 + \dd\varphi^2) \,,
    \label{conformal-metric-simplification-2}
\end{eqnarray}
so  that the conformal metric $h_{ab}$ on scri takes a simple form
\begin{equation}
h = \dfrac{\Lambda}{3}\,
   \Big[(A_\scrisub\dd t - B_\scrisub\dd\varphi)^2
   - \frac{(A^2_\scrisub + \alpha^2 B^2_\scrisub)^2}{{P}_\scrisub {Q}_\scrisub}\,\dd x^2 \Big]
   + {P}_\scrisub(\dd\varphi^2 + \alpha^2 \dd t^2 ) \,.
    \label{conformal-metric-final}
\end{equation}

It can be used do determine the presence of gravitational radiation generated by accelerating black holes in the entire type D family. Notice that the metric depends only on the coordinate~$x$ via the functions ${A_\scrisub=1 + \alpha^2(l^2-a^2)\,x^2}$, ${B_\scrisub=a + 2l\,x + a\,x^2}$, and ${P}_\scrisub(x)$, ${Q}_\scrisub(x)$ given by \eqref{P_on_scri}, \eqref{Q_on_scri}. It is \emph{everywhere regular}, except at ${{P}_\scrisub=0}$ and/or ${{Q}_\scrisub=0}$ where the axes of symmetry and/or the horizons intersect the conformal infinity.

All subclasses of black holes can easily be obtained from \eqref{conformal-metric-final} by setting the corresponding physical parameter(s) to zero. Interestingly, the dependence on the mass and charges $m, e, g$ is \emph{only implicit} via the metric functions \eqref{P_on_scri}, \eqref{Q_on_scri} because they do not enter $A_\scrisub$ nor $B_\scrisub$.

Big simplification occurs in the case of \emph{non-accelerating} black holes. Indeed, for ${\alpha=0}$ the metric $h_{ab}$, corresponding to the Kerr-Newman-NUT-(A)dS black hole, reduces to
\begin{align}
h = & \ \dfrac{\Lambda}{3}\big(\dd t - [\,a\,(1 + x^2) + 2l\,x\,]\,\dd\varphi \big)^2  + \frac{\dd x^2 }{P_\scrisub}    + P_\scrisub \dd\varphi^2 \,,
    \label{conformal-metric-final-alpha=0}\\
&\hbox{where}\qquad
P_\scrisub = (1-x^2) - \frac{\Lambda}{3} (l+a\,x)(3l+a\,x)\,x^2 . \label{P_on_scri-alpha=0}
\end{align}
For ${a=0=l}$, with ${x=\cos\theta}$, we obtain the Schwarzschild-(A)dS metric on $\scri$, namely
\begin{align}
h = \frac{\Lambda}{3}\,\dd t^2 +\dd\theta^2 + \sin^2\theta\,\dd\varphi^2 \, .
    \label{Schw-AdS-metric}
\end{align}
When ${\Lambda=0}$, it is just a 2-sphere, as expected. For ${\Lambda>0}$ it is a Riemannian 3-space, while for ${\Lambda<0}$ it is a Lorentzian (1+2)-space.

Complementarily, for the \emph{simplest accelerating} black holes without twist and charge (${a=0=l}$, ${e=0=g}$), that is the C-metric in (A)dS, the metric  \eqref{conformal-metric-final} on scri reduces~to
\begin{equation}
h =  - {Q}_\scrisub \dd t^2
     - \dfrac{\Lambda}{3}\,\frac{\dd x^2}{{P}_\scrisub {Q}_\scrisub}
    + {P}_\scrisub \dd\varphi^2  \,,
    \label{conformal-C-metric}
\end{equation}
with ${{P}_\scrisub = (1-x^2)(1-2\alpha m \,x)}$, and
${-{Q}_\scrisub = \frac{1}{3}\Lambda + \alpha^2{P}_\scrisub}$ (so that ${\rho^2_\scrisub = 1}$).
It follows from  \eqref{conformal-C-metric} that ${{Q}_\scrisub=0}$ for ${\Lambda<0}$ identifies the Killing horizon of this conformal metric, with the null generator $\partial_{t}$.

\subsection{The normal to $\scri$}
\label{subsectionc:normal-to-scri}

Finally, we will determine the {\em causal character} of~$\scri$. Consider the  \emph{normal} ${\bf N}$ to the hypersurfaces ${\Omega=\hbox{const.}}$ defined by ${N^\alpha \defeq g^{\alpha\beta}N_\beta}$, where
\begin{equation}
N_\beta \defeq \nabla_\beta\,\Omega = \dd_\beta\,\Omega = \Omega_{,\beta}\,.
    \label{norm-N_alpha}
\end{equation}
Since the differential of ${\,\Omega\,}$ given by \eqref{Omega_new} is simply
\begin{align}
\dd\Omega = \dd q - \alpha\, \dd x\,,
    \label{d Omega}
\end{align}
and the conformal (unphysical) metric is \eqref{conformal-metricJP2024}, we get
	\begin{equation}
	{\bf N} = \frac{1}{\rho^2}\,\big( {Q} \,\partial_q - \alpha\,{P}\,\partial_x \big) \,.
	\label{eq:normal-scri}
	\end{equation}
Its \emph{norm} is ${g_{\alpha\beta}\,N^\alpha N^\beta=( {Q} + \alpha^2 {P})/\rho^{2}}$, and applying the identity (\ref{Q-on-scri-short}) valid on $ \scri $ we get
	\begin{equation}
	\big(\,{g}_{\alpha\beta}\,{N}^{\alpha}{N}^{\beta}\big)_\scrisub = -\frac{\Lambda}{3}\, . \label{norm-ofn--on-scri}
	\end{equation}
The causal character of $\scri$ is thus fully determined just by the sign of the cosmological constant~$\Lambda$: it is \emph{null} for ${\Lambda=0}$, \emph{spacelike} for ${\Lambda>0}$, and \emph{timelike} for ${\Lambda<0}$.

We will also need an explicit \emph{unit normal $\mathbf{n}$}~to $\scri$ defined as
\begin{equation}	\label{eq:unit-normal-scri}
 \mathbf{n}\defeq\frac{1}{\sqrt{|g_{\alpha\beta}\,N^\alpha N^\beta|}}\,\mathbf{N}
       =\sqrt{\frac{3}{|\Lambda|}}\,\mathbf{N}\,,
\end{equation}
that is
\begin{equation}	\label{eq:unit-normal-scri-explicit}
 \mathbf{n} =\sqrt{\frac{3}{|\Lambda|}}\,\,\frac{1}{\rho^2_\scrisub}
     \big( {Q}_\scrisub \partial_q - \alpha\,{P}_\scrisub\partial_x \big) \,.
\end{equation}
For ${\Lambda>0}$ it is a (future-pointing) \emph{unit timelike normal} to de~Sitter-like~$\scri$, while for ${\Lambda<0}$ it is a \emph{unit spacelike normal} to anti-de~Sitter-like~$\scri$ (oriented inwards or outwards for ${Q_\scrisub>0}$ or ${Q_\scrisub<0}$, respectively). Indeed, using the metric \eqref{conformal-metricJP2024} and identity  \eqref{Q-on-scri-short} we immediately derive that
${\mathbf{n}\cdot\mathbf{n}\equiv g_{\alpha\beta}\,n^\alpha n^\beta = -\hbox{sign}\,\Lambda}$.

This global structure is also reflected in the conformal metric $h_{ab}$ on $\scri$ given by  \eqref{conformal-metric-final}. In it, ${{P}_\scrisub > 0 }$ everywhere (with ${{P}_\scrisub = 0 }$ identifying the axes of symmetry), and the 1-forms are quadratic. Therefore, the signature of $h_{ab}$ depends only on the sign of the cosmological constant $\Lambda$, and on the sign of ${Q}_\scrisub$. Obviously, \emph{for} ${\Lambda>0}$ the metric $h$ given by \eqref{conformal-metric-final} is \emph{Riemannian} with the signature ${(+,+,+)}$ because ${Q<0}$ close to the de~Sitter-like $\scri$ (above the cosmological horizon).
On the other hand, \emph{for} ${\Lambda<0}$ the metric is \emph{Lorentzian} on the anti-de~Sitter-like $\scri$, with the signature ${(-,+,+)}$. Actually, the coordinate~$x$ is spacelike if ${Q_\scrisub >0 }$, while it is timelike if ${Q_\scrisub <0 }$. Since ${Q_\scrisub =0}$ separates these regions in which~$x$ has different causal character, in this sense it can be understood as the ``intersection of a horizon with~$\scri$''.

\section{The PND frame in the case ${\Lambda>0}$}
\label{section:PND-Weyl-for-Lambda>0}

To determine covariant quantities which characterize the gravitational radiation at~$\scri$, it is necessary to find an \emph{appropriate null tetrad} adapted to principal null directions (PNDs) of the Weyl curvature tensor. It turns out that in the non-stationary region near the de~Sitter-like conformal infinity~$\scri$ where ${Q<0}$ such a tetrad in the physical metric $\hat{g}_{\alpha\beta}$ has the form
\begin{eqnarray}
	\mathbf{\hat{k}} \rovno \frac{1}{\sqrt{2}}\, \frac{\Omega}{\rho}
        \bigg[ \frac{1}{\sqrt{-{Q}}}
	 \,\textbf{T} + \sqrt{-{Q}}\, \partial_q \bigg] , \nonumber \\[2mm]
	\mathbf{\hat{l}} \rovno \frac{1}{\sqrt{2}}\, \frac{\Omega}{\rho}
        \bigg[ \frac{-1}{\sqrt{-{Q}}}
	 \,\textbf{T} + \sqrt{-{Q}}\, \partial_q \bigg] ,  \label{nullframe-Q<0}\\[2mm]
	\mathbf{\hat{m}} \rovno \frac{1}{\sqrt{2}}\, \frac{\Omega}{\rho} \bigg[
	 \frac{1}{\sqrt{{P}}} \,\textbf{R} - \mathrm{i} \, \sqrt{{P}}\, \partial_x \bigg] , \nonumber
\end{eqnarray}
where the Killing vector fields $\textbf{T}$ and $\textbf{R}$ are defined as
\begin{align}
	\textbf{T} &\defeq  D \, \partial_{t} - C \, \partial_{\varphi}  \,, \nonumber\\
	\textbf{R} &\defeq  B \, \partial_{t} + A \, \partial_{\varphi}  \,.  \label{TRdef}
\end{align}
Using the physical metric \eqref{A-new-metric} and the identity \eqref{rho2_new} it can be shown  that
\begin{eqnarray}
	\textbf{T}\cdot\textbf{T} = -{Q} \,\frac{\rho^2}{\Omega^2} \,,\qquad
	\textbf{R}\cdot\textbf{R} =  {P} \,\frac{\rho^2}{\Omega^2} \,,\qquad
	\textbf{T}\cdot\textbf{R} =  0 \,,  \label{TTTTR}
\end{eqnarray}
so that  the vector field $\textbf{T}$ is \emph{timelike} in the regions ${Q>0}$, \emph{spacelike} in the regions ${Q<0}$, and it is \emph{null} on the (Killing) horizons where ${Q=0}$. On the other hand, the vector field $\textbf{R}$ is \emph{everywhere spacelike}, except at ${P=0}$ where its norm vanishes, defining thus geometrically the axes of axial symmetry. Moreover, these vector fields $\textbf{T}$  and $\textbf{R}$ are \emph{mutually orthogonal}. For the static case ${a=0=l}$ we get simply ${\textbf{T} = \partial_{t}}$ and ${\textbf{R} = \partial_{\varphi}}$, i.e., the usual generators of time translation isometry (for ${Q>0}$) and of the axial symmetry.

Applying the relations \eqref{TTTTR} it is easy to check that the tetrad \eqref{nullframe-Q<0} satisfies the normalization conditions ${\mathbf{\hat{k}}\cdot\mathbf{\hat{l}}=-1}$ and ${\mathbf{\hat{m}}\cdot\mathbf{\bar{\hat{m}}}=1}$ (with all other scalar products vanishing). Moreover, it is adapted to \emph{both  double-repeated PNDs $\mathbf{\hat{k}}$ and $\mathbf{\hat{l}}$}. This is confirmed by evaluating the components of the Weyl tensor with respect to this tetrad, that is the Newman-Penrose scalars. It turns out that the only non-vanishing such scalar is~$\Psi_2$.

The null tetrad \eqref{nullframe-Q<0} is directly associated to a \emph{special orthonormal tetrad} ${(\mathbf{\hat{t}}, \mathbf{\hat{q}}, \mathbf{\hat{r}}, \mathbf{\hat{s}})}$ via the usual algebraic relations
\begin{equation}
	\mathbf{\hat{k}} \defeq  \tfrac{1}{\sqrt2} (\mathbf{\hat{t}} + \mathbf{\hat{q}})\,, \qquad
	\mathbf{\hat{l}} \defeq  \tfrac{1}{\sqrt2} (\mathbf{\hat{t}} - \mathbf{\hat{q}})\,, \qquad
	\mathbf{\hat{m}} \defeq  \tfrac{1}{\sqrt2} (\mathbf{\hat{s}} + \mathrm{i}\,
    \mathbf{\hat{r}})\,,  \label{special-orthonormal-def}
\end{equation}
that is explicitly
\begin{eqnarray}
	 \mathbf{\hat{t}} \rovno  \frac{\Omega}{\rho} \,\sqrt{-{Q}}\, \partial_{q} \,, \qquad
	 \mathbf{\hat{q}} =  \frac{\Omega}{\rho}\,\frac{1}{\sqrt{-{Q}}}\,\, \textbf{T} \,,  \nonumber\\
	 \mathbf{\hat{r}} \rovno -\frac{\Omega}{\rho}\,\sqrt{{P}}\, \partial_x \,,  \qquad
	 \mathbf{\hat{s}} = \frac{\Omega}{\rho}\,\frac{1}{\sqrt{{P}}}\,\, \textbf{R} \,.
    \label{special-orthonormal}
\end{eqnarray}
For the metric \eqref{A-new-metric} these vectors are normalized  as ${\mathbf{\hat{t}}\cdot\mathbf{\hat{t}}=-1}$ and ${\mathbf{\hat{q}}\cdot\mathbf{\hat{q}}=\mathbf{\hat{r}}\cdot\mathbf{\hat{r}}=\mathbf{\hat{s}}\cdot\mathbf{\hat{s}}=1}$ (all other scalar products vanish).

In the following, we will need the related null tetrad ${(\mathbf{k},\mathbf{l},\mathbf{m})}$ on $\scri$ in the unphysical metric $g_{\alpha\beta}$ that is future-oriented and aligned with the two double PNDs $\mathbf{\hat{k}}$ and $\mathbf{\hat{l}}$. This is easily obtained from \eqref{nullframe-Q<0}, by rescaling it by $\Omega^{-1}$, and inverting the orientation of $\mathbf{k} $ and $\mathbf{l}$:
\begin{eqnarray}
	\mathbf{k} \rovno -\frac{1}{\sqrt{2}}\, \frac{1}{\rho} \bigg[ \frac{1}{\sqrt{-{Q}}}
	 \,\textbf{T} + \sqrt{-{Q}}\, \partial_q \bigg] , \nonumber \\[2mm]
	\mathbf{l} \rovno -\frac{1}{\sqrt{2}}\, \frac{1}{\rho} \bigg[ \frac{-1}{\sqrt{-{Q}}}
	 \,\textbf{T} + \sqrt{-{Q}}\, \partial_q \bigg] ,  \label{nulltetrad-Q<0}\\[2mm]
	\mathbf{m} \rovno \frac{1}{\sqrt{2}}\, \frac{1}{\rho} \bigg[
	 \frac{1}{\sqrt{{P}}} \,\textbf{R} - \mathrm{i} \, \sqrt{{P}}\, \partial_x \bigg] . \nonumber
\end{eqnarray}
It can be checked using the unphysical metric \eqref{conformal-metricJP2024} that that the corresponding orthonormal tetrad
\begin{eqnarray}
	 \mathbf{t} \rovno  -\frac{\sqrt{-{Q}} }{\rho}\, \partial_{q} \,, \qquad
	 \mathbf{q} =  -\frac{1}{\rho\,\sqrt{-Q}}\,\, \textbf{T}\,,  \nonumber\\
	 \mathbf{r} \rovno -\frac{\sqrt{{P}}}{\rho}\, \partial_x  \,,  \qquad
	 \mathbf{s} = \frac{1}{\rho\,\sqrt{P}}\,\, \textbf{R} \,,
    \label{Q<0-orthonormal}
\end{eqnarray}
satisfies ${\mathbf{t}\cdot\mathbf{t}=-1}$ and ${\mathbf{q}\cdot\mathbf{q}=\mathbf{r}\cdot\mathbf{r}=\mathbf{s}\cdot\mathbf{s}=1}$. Their values on $\scri$ are obtained by employing the functions ${P_\scrisub, Q_\scrisub}$ and $A_\scrisub, B_\scrisub, C_\scrisub, D_\scrisub$ (entering ${\textbf{T}, \textbf{R}}$) given by \eqref{AB_on_scri}--\eqref{Q_on_scri}.

\section{The PND frames in the case ${\Lambda<0}$}
\label{section:PND-Weyl-for-Lambda<0}

For ${\Lambda<0}$, a new interesting feature appears. Depending on the physical parameters of the black hole, regions near the corresponding \emph{anti-de~Sitter-like} conformal infinity~$\scri$ can be either \emph{stationary} (${{Q}>0}$) or \emph{non-stationary} (${{Q}<0}$), and they are ``separated by horizons'' (${{Q}=0}$).

\subsection{The structure of scri and the unit normal}
\label{subsectionc:Lambda<0}

This fact follows from the crucial identity  \eqref{Q-on-scri-short} on~$\scri$, which using \eqref{rho2_on_scri} can be written as
\begin{eqnarray}
 {Q}_\scrisub = -\frac{\Lambda}{3}\,(A_\scrisub^2 + \alpha^2 B_\scrisub^2)
 - \alpha^2 {P}_\scrisub  \,. \label{Q-on-scri-alternative}
\end{eqnarray}
For ${\Lambda>0}$, the right-hand side is always negative because ${P_\scrisub\ge0}$ everywhere, to keep the correct signature of the metric \eqref{A-new-metric}, and thus ${Q_\scrisub <0}$ everywhere on~$\scri$, as discussed in Sec.~\ref{subsectionc:normal-to-scri}. But for ${\Lambda<0}$ it is necessary to distinguish three distinct possibilities, namely ${Q_\scrisub >0}$, ${Q_\scrisub <0}$, and ${Q_\scrisub =0}$. These conditions identify \emph{three different types of regions} of the anti-de~Sitter scri, whose normal is a \emph{spatial} unit vector \eqref{eq:unit-normal-scri-explicit}.

In fact, for \emph{zero or small acceleration} of the black hole, the second term on the right-hand side of  \eqref{Q-on-scri-alternative}  is negligible with respect to the first term (which is positive), so that ${Q_\scrisub >0}$ \emph{everywhere} on~$\scri$. On the other hand, for \emph{large acceleration} there can be ${Q_\scrisub <0}$ somewhere on~$\scri$. These two regions of~$\scri$ are separated by the boundary ${Q_\scrisub =0}$, that is located at
\begin{eqnarray}
 \alpha^2 {P}_\scrisub  = -\frac{\Lambda}{3}\,(A_\scrisub^2 + \alpha^2 B_\scrisub^2 )\,. \label{Q-on-scri-alternative2}
\end{eqnarray}
This is \emph{formally} the \emph{``intersection'' of the horizons with the anti-de~Sitter-like scri}. Substituting from \eqref{AB_on_scri} and \eqref{P_on_scri} we obtain the explicit 4th-order polynomial equation in $x$.

In particular, in the ``middle plane'' ${x=0}$ there is ${{P}_\scrisub = 1+\alpha^2a^2}$, ${A_\scrisub = 1}$, ${B_\scrisub  = a}$, so that the condition \eqref{Q-on-scri-alternative2} for ${{Q}_\scrisub =0}$ reduces to a simple relation
\begin{eqnarray}
 \alpha^2 = -\frac{\Lambda}{3} \,. \label{Q-on-scri-for-x=0}
\end{eqnarray}
For \emph{small acceleration} ${\alpha < \sqrt{-\Lambda/3}}$ it follows from \eqref{Q-on-scri-alternative} that ${Q_\scrisub >0}$ which means that the region close to the section ${x=0}$ of $\scri$ is \emph{stationary}, whereas for \emph{large acceleration} ${\alpha > \sqrt{-\Lambda/3}}$  such a region close to ${x=0}$ on $\scri$ is \emph{non-stationary} since  ${Q_\scrisub <0}$. This has been previously observed for (non-twisting) C-metric with ${\Lambda<0}$, i.e., for black holes accelerating in the anti-de~Sitter universe \cite{Podolsky:2002, PodolskyOrtaggioKrtous:2003}. Actually the ${\alpha < \sqrt{-\Lambda/3}}$ case represents a \emph{single} accelerated black holes, while the ${\alpha > \sqrt{-\Lambda/3}}$ case represents \emph{a pair} of accelerated black holes, see in particular Fig.~9 and Fig.~10 in \cite{PodolskyOrtaggioKrtous:2003}.

Therefore, for the full analysis we must now separately discuss the three possibilities, namely ${Q_\scrisub >0}$, ${Q_\scrisub <0}$, and ${Q_\scrisub =0}$. To this end, it is useful to recall that the unphysical (conformal) metric~\eqref{conformal-metricJP2024} is
\begin{equation}
    \dd s^2 = -\dfrac{{Q}}{\rho^2}(A\,\dd t - B\,\dd\varphi)^2
    + \dfrac{{P}}{\rho^2}(C\,\dd t + D\,\dd\varphi)^2
    + \rho^2\Big(\,\dfrac{\dd q^2}{{Q}} + \dfrac{\dd x^2}{{P}}\,\Big)\,.
    \label{conformal-metric}
\end{equation}
The corresponding anti-de~Sitter-like scri~$\scri$ is identified by ${\Omega=0}$, with the 1-form ${\dd\Omega = \dd q - \alpha\, \dd x}$, see~\eqref{d Omega}. It has the normal vector ${N^\alpha = g^{\alpha\beta}\,(\dd\,\Omega)_\beta = g^{\alpha\beta}\,\Omega_{,\beta}}$, and it gives rise to the \emph{unit spacelike} normal~$\mathbf{n}$ to~$\scri$, cf. \eqref{eq:unit-normal-scri}, \eqref{eq:unit-normal-scri-explicit},
\begin{equation}	\label{eq:unit-normal-scri-Lambdsa<0}
 \mathbf{n} =\sqrt{-\frac{3}{\Lambda}}\,\frac{1}{\rho^2_\scrisub}\,
     \big( {Q}_\scrisub \partial_q - \alpha\,{P}_\scrisub\partial_x \big) \,.
\end{equation}
such that ${\mathbf{n}\cdot\mathbf{n}=1}$. It is oriented inwards (towards the physical spacetime) when ${Q_\scrisub>0}$, and outwards (out of it) when ${Q_\scrisub<0}$.

Moreover, with the metric  \eqref{conformal-metric} the vectors
${\textbf{T} =  D \, \partial_{t} - C \, \partial_{\varphi}}$ and
${\textbf{R} =  B \, \partial_{t} + A \, \partial_{\varphi}}$, introduced in \eqref{TRdef}, now evaluated on scri using the relations \eqref{AB_on_scri}--\eqref{rho2_on_scri}, satisfy
\begin{equation}
	(\textbf{T}\cdot\textbf{T})_\scrisub = -{Q}_\scrisub \,\rho^2_\scrisub \,,\qquad
	(\textbf{R}\cdot\textbf{R})_\scrisub =  {P}_\scrisub \,\rho^2_\scrisub \,,\qquad
	(\textbf{T}\cdot\textbf{R})_\scrisub =  0 \,.  \label{TTTTR on scri}
\end{equation}
It implies that the field $\textbf{R}$ is \emph{spacelike everywhere} on~$\scri$. On the other hand, the field $\textbf{T}$ is \emph{timelike} in the region ${Q_\scrisub>0}$, \emph{spacelike} in the region ${Q_\scrisub<0}$, and it is \emph{null} on the ``horizon boundary'' where ${Q_\scrisub=0}$. (Interestingly, $\textbf{T}$ and $\textbf{R}$ are always mutually orthogonal.)

Recall that in the static case ${a=0=l}$ the coefficients reduce to ${A_\scrisub=1=D_\scrisub}$, ${B_\scrisub=0=C_\scrisub}$, so that ${\textbf{T}=\partial_{t}}$ and ${\textbf{R}=\partial_{\varphi}}$ are the usual \emph{Killing vector fields}. In the general case, however the coefficients $A_\scrisub, B_\scrisub, C_\scrisub, D_\scrisub$ are functions of~$x$, and thus $\textbf{T}$ and $\textbf{R}$ are \emph{not} the generators of symmetries in the spacetime \eqref{conformal-metric-final}.

\newpage

\subsection{The PND frame in stationary regions ${Q_\scrisub >0}$}
\label{subsectionc:case-Q>0}

In the stationary regions of a ${\Lambda<0}$ spacetime, close to the anti-de~Sitter~$\scri$ with ${Q_\scrisub >0}$, we will apply the following  null tetrad adapted to the PNDs of the Weyl curvature tensor,
\begin{eqnarray}
	\mathbf{\hat{k}} \rovno \frac{1}{\sqrt{2}}\, \frac{\Omega}{\rho}
        \bigg[ \frac{1}{\sqrt{{Q}}}
	 \,\textbf{T} - \sqrt{{Q}}\, \partial_q \bigg] , \nonumber \\[2mm]
	\mathbf{\hat{l}} \rovno \frac{1}{\sqrt{2}}\, \frac{\Omega}{\rho}
        \bigg[ \frac{1}{\sqrt{{Q}}}
	 \,\textbf{T} + \sqrt{{Q}}\, \partial_q \bigg] ,  \label{nullframe-Q>0physical}\\[2mm]
	\mathbf{\hat{m}} \rovno \frac{1}{\sqrt{2}}\, \frac{\Omega}{\rho} \bigg[
	 \frac{1}{\sqrt{{P}}} \,\textbf{R} - \mathrm{i} \, \sqrt{{P}}\, \partial_x \bigg] . \nonumber
\end{eqnarray}
Here $\textbf{T}$ is the specific (future-oriented) \emph{timelike} vector field \eqref{TRdef} on~$\scri$, see  \eqref{TTTTR on scri}, while $\partial_q$ is the \emph{spacelike} vector field (because ${g_{qq}>0}$). This null tetrad is related via \eqref{special-orthonormal-def} to a special orthonormal tetrad in the physical spacetime $\hat{g}_{\alpha\beta}$:
\begin{eqnarray}
	 \mathbf{\hat{t}} \rovno \frac{\Omega}{\rho}\frac{1}{\sqrt{Q}}\,\, \textbf{T} \,, \qquad
	 \mathbf{\hat{q}} = -\frac{\Omega}{\rho}\,\sqrt{Q}\, \partial_{q}\,, \nonumber\\
	 \mathbf{\hat{r}} \rovno -\frac{\Omega}{\rho}\,\sqrt{{P}}\, \partial_x \,,  \qquad
	 \mathbf{\hat{s}} = \frac{\Omega}{\rho} \,\frac{1}{\sqrt{P}}\,\, \textbf{R} \,,
    \label{Q>0-orthonormalphysical}
\end{eqnarray}
normalized  as ${\mathbf{\hat{t}}\cdot\mathbf{\hat{t}}=-1}$ and ${\mathbf{\hat{q}}\cdot\mathbf{\hat{q}}=\mathbf{\hat{r}}\cdot\mathbf{\hat{r}}=\mathbf{\hat{s}}\cdot\mathbf{\hat{s}}=1}$
with the metric \eqref{A-new-metric}.

The related future-oriented null tetrad ${(\mathbf{k},\mathbf{l},\mathbf{m})}$ in the conformally related unphysical spacetime $g_{\alpha\beta}$ is obtained by rescaling it by $\Omega^{-1}$:
\begin{eqnarray}
	\mathbf{k} \rovno \frac{1}{\sqrt{2}}\, \frac{1}{\rho} \bigg[ \frac{1}{\sqrt{Q}}
	 \,\textbf{T} - \sqrt{{Q}} \, \partial_q \bigg] , \nonumber \\[2mm]
	\mathbf{l} \rovno \frac{1}{\sqrt{2}}\, \frac{1}{\rho} \bigg[ \frac{1}{\sqrt{Q}}
	 \,\textbf{T} + \sqrt{{Q}} \, \partial_q \bigg] ,  \label{nullframe-Q>0}\\[2mm]
	\mathbf{m} \rovno \frac{1}{\sqrt{2}}\, \frac{1}{\rho} \bigg[
	 \frac{1}{\sqrt{P}} \,\textbf{R} - \mathrm{i} \, \sqrt{{P}}\, \partial_x \bigg] . \nonumber
\end{eqnarray}
Using ${\dd\Omega = \dd q - \alpha\, \dd x}$, it is easy to evaluate the contractions on $\scri$:
\begin{equation}	\label{eq:PND-orientations-Q>0}
 -\langle \dd\Omega , \mathbf{k}   \rangle_\scrisub =  \langle \dd\Omega , \mathbf{l}   \rangle_\scrisub
  = \frac{\sqrt{{Q}_\scrisub}}{\sqrt 2\,\rho_\scrisub}\,.
\end{equation}
Because ${\langle \dd\Omega , \mathbf{k}   \rangle_\scrisub <0}$ while ${\langle \dd\Omega , \mathbf{l}   \rangle_\scrisub >0}$, we conclude that the two principal null directions are oriented \emph{oppositely} on~$\scri$. Namely, the PND $\mathbf{k}$ \emph{points outwards}, while the PND $\mathbf{l}$ \emph{points inwards}. (By simply changing $\partial_q$ to $-\partial_q$ in \eqref{nullframe-Q>0} we get exactly the opposite orientations.)


It can also be observed from \eqref{eq:PND-orientations-Q>0} that ``on the horizon'' where ${Q_\scrisub =0}$, \emph{both the PND vectors $\mathbf{k}$ and $\mathbf{l}$ are tangent to~$\scri$.} Actually, by a suitable rescaling (boost) we can obtain ${\mathbf{k}_{\tiny\hbox{horizon}}\propto \textbf{T}}$ and, independently, ${\mathbf{l}{\tiny\hbox{horizon}}\propto \textbf{T}}$. This is consistent with the fact that ${(\textbf{T}\cdot\textbf{T})_\scrisub = 0}$ on the intersection of horizon with the scri, see \eqref{TTTTR on scri}.

The corresponding orthonormal tetrad adapted to the geometry of PNDs is
\begin{eqnarray}
	 \mathbf{t} \rovno  \frac{1}{\rho\,\sqrt{Q}}\,\, \textbf{T}  \,, \qquad
	 \mathbf{q} =  -\frac{\sqrt{{Q}} }{\rho}\, \partial_{q}\,,  \nonumber\\
	 \mathbf{r} \rovno -\frac{\sqrt{{P}} }{\rho}\, \partial_x \,,  \qquad
	 \mathbf{s} = \frac{1}{\rho\,\sqrt{P}}\,\, \textbf{R} \,.
    \label{Q>0-orthonormal}
\end{eqnarray}
Using the unphysical metric \eqref{conformal-metric} and the relations \eqref{TTTTR on scri} on scri we immediately get ${\mathbf{t}\cdot\mathbf{t}=-1}$ and ${\mathbf{q}\cdot\mathbf{q}=\mathbf{r}\cdot\mathbf{r}=\mathbf{s}\cdot\mathbf{s}=1}$ (all other scalar products vanish). Moreover, the \emph{spacelike normal}~$\mathbf{n}$ to~$\scri$, given by \eqref{eq:unit-normal-scri-Lambdsa<0} such that ${\mathbf{n}\cdot\mathbf{n}=1}$, expressed in this tetrad reads
\begin{equation}	\label{eq:unit-normal-Q>0}
 \mathbf{n} =\sqrt{-\frac{3}{\Lambda}}\,\,\frac{1}{\rho_\scrisub}\,
     \,\Big( -\sqrt{{Q}_\scrisub}\, \mathbf{q} + \alpha\,\sqrt{{P}_\scrisub}\,\mathbf{r} \Big) \,.
\end{equation}

\subsection{The PND frame in non-stationary regions ${Q_\scrisub <0}$}
\label{subsectionc:case-Q<0}

In this region, we meet a complementary situation. In particular, we can employ the PND frame \eqref{nulltetrad-Q<0}, that is
\begin{eqnarray}
	\mathbf{k} \rovno -\frac{1}{\sqrt{2}}\, \frac{1}{\rho} \bigg[ \frac{1}{\sqrt{-Q}}
	 \,\textbf{T} + \sqrt{-{Q}}\,\partial_q \bigg] , \nonumber \\[2mm]
	\mathbf{l} \rovno -\frac{1}{\sqrt{2}}\, \frac{1}{\rho} \bigg[ \frac{-1}{\sqrt{-Q}}
	 \,\textbf{T} + \sqrt{-{Q}}\,\partial_q \bigg] ,  \label{nulltetrad-Q<0-again}\\[2mm]
	\mathbf{m} \rovno \frac{1}{\sqrt{2}}\, \frac{1}{\rho} \bigg[
	 \frac{1}{\sqrt{P}} \,\textbf{R} - \mathrm{i} \,\sqrt{{P}}\,\partial_x \bigg] . \nonumber
\end{eqnarray}
However, it follows from \eqref{TTTTR on scri} that $\textbf{T}$ is now \emph{spacelike} on $\scri$. On the other hand,  $\partial_q$ is the \emph{timelike} vector field on~$\scri$ because ${g_{qq}<0}$ in  the metric \eqref{conformal-metric}.
Contractions of $\dd\Omega$ with $\mathbf{k}$ and $\mathbf{l}$ give
\begin{equation}	\label{eq:PND-orientations-Q<0}
 \langle \dd\Omega , \mathbf{k}   \rangle_\scrisub =  \langle \dd\Omega , \mathbf{l}   \rangle_\scrisub
  = -\frac{\sqrt{-{Q}_\scrisub}}{\sqrt 2\,\rho_\scrisub} \,.
\end{equation}
Because ${\langle \dd\Omega , \mathbf{k}   \rangle_\scrisub <0}$ and also ${\langle \dd\Omega , \mathbf{l}   \rangle_\scrisub <0}$, both  the principal null directions are oriented \emph{in the same direction} on~$\scri$. Namely, \emph{both~$\mathbf{k}$ and~$\mathbf{l}$ point outwards} (this corresponds to the regions denoted as ${{\cal I}_{{\rm I}^+}}$ in Fig.~10 of~\cite{PodolskyOrtaggioKrtous:2003}). If we wish to study regions of $\scri$ where \emph{both~$\mathbf{k}$ and~$\mathbf{l}$ point inwards} (the regions denoted as ${{\cal I}_{{\rm I}^-}}$ in~\cite{PodolskyOrtaggioKrtous:2003}), we take the analogue of \eqref{nulltetrad-Q<0-again} in which we change ${\partial_q}$ to ${-\partial_q}$.

\newpage

The related orthonormal tetrad is
\begin{eqnarray}
	 \mathbf{t} \rovno -\frac{\sqrt{-{Q}} }{\rho}\, \partial_{q} \,, \qquad
	 \mathbf{q} =  -\frac{1}{\rho\,\sqrt{-Q}}\,\, \textbf{T}\,,  \nonumber\\
	 \mathbf{r} \rovno -\frac{\sqrt{{P}}}{\rho}\, \partial_x \,,  \qquad
	 \mathbf{s} = \frac{1}{\rho\,\sqrt{P}}\,\, \textbf{R}  \,.
    \label{Q<0-orthonormal2}
\end{eqnarray}
On  $\scri$, such $\mathbf{t}$ is timelike, and it is future-oriented for $q$~decreasing (when approaching~$\scri$), and it is past-oriented for $q$~increasing (when receding from~$\scri$).

The spacelike unit normal~$\mathbf{n}$ to~$\scri$, given by \eqref{eq:unit-normal-scri-Lambdsa<0}, now takes the form
\begin{equation}	\label{eq:unit-normal-Q<0}
 \mathbf{n} =\sqrt{-\frac{3}{\Lambda}}\,\,\frac{1}{\rho_\scrisub}\,
     \,\Big( \sqrt{-{Q}_\scrisub}\, \mathbf{t} + \alpha\,\sqrt{{P}_\scrisub}\,\mathbf{r} \Big) \,.
\end{equation}
This may seem like an inconsistency, namely that  $\mathbf{t} $ is timelike while $\mathbf{n}$ is spacelike. But we should emphasize that \emph{here it is not possible to consider small values of the acceleration parameter~$\alpha$}, or even to take ${\alpha=0}$. Indeed, the  non-stationary region of~$\scri$ with ${Q_\scrisub <0}$, which we are considering, here occurs only if
\begin{eqnarray}
\alpha^2 {P}_\scrisub  >
 -\frac{\Lambda}{3}\,(A_\scrisub^2 + \alpha^2 B_\scrisub^2) > 0 \,, \label{Q<0 condition}
\end{eqnarray}
see \eqref{Q-on-scri-alternative}. Small values of~$\alpha$ are thus not allowed.

\subsection{The case ${Q_\scrisub=0}$}
\label{subsectionc:case-Q=0}

In both the stationary case and the non-stationary case, by setting ${Q_\scrisub =0}$ in \eqref{eq:PND-orientations-Q>0} and \eqref{eq:PND-orientations-Q<0} we immediately get
\begin{equation}	\label{eq:PND-orientations-Q=0}
 \langle \dd\Omega , \mathbf{k}   \rangle_\scrisub = 0 =  \langle \dd\Omega , \mathbf{l}   \rangle_\scrisub
 \,.
\end{equation}
It means that both the PNDs $\mathbf{k} $ and $\mathbf{l} $ are \emph{tangent to}~$\scri$ whenever ${Q_\scrisub =0}$. Since ${Q=0}$ determines the position of the Killing horizons in the physical spacetime, we can naturally interpret ${Q_\scrisub =0}$ as the  ``horizon boundary''  at~$\scri$ that separates the stationary regions ${Q_\scrisub>0}$ from the non-stationary regions ${Q_\scrisub<0}$. The fact that both PNDs are tangent to the ${\Lambda<0}$ scri at ${Q_\scrisub=0}$ was observed in the C-metric in anti-de~Sitter universe already in the work \cite{PodolskyOrtaggioKrtous:2003}, and subsequently in a general setting analyzed in the review article \cite{KrtousPodolsky:2004}, see in particular Table~1 therein.

\newpage

\section{Study of the gravitational radiation}

Now we will apply the new physical metric \eqref{A-new-metric} and the geometrical objects on its conformal infinity~$\scri$ (namely the canonical frames and their relations to the normal, presented in Sec.~\ref{sections:scri-and-hab}--Sec.~\ref{section:PND-Weyl-for-Lambda<0}) to the study of gravitational radiation in the whole Pleba\'nski-Demia\'nski class of black holes with \emph{any} value of the cosmological constant~$\Lambda\ne0$.

This investigation has been performed in collaboration with my colleagues Francisco Fern\'andez-\'Alvarez and Jos\'e~M.~M.~Senovilla, that involved many discussions and mutual feedbacks. In fact, it is an extension of our previous work \cite{FernandezPodolskySenovilla:2024}, in which we analyzed gravitational radiation generated by the type D black holes with a positive cosmological constant, to a more complicated case of a \emph{negative} cosmological constant. This study is based on the works of Francisco and Jos\'e, namely their recent paper \cite{FernandezSenovilla:2026}. Another key improvement is the application of the complete family of Pleba\'nski-Demia\'nski black holes in the new form \eqref{A-new-metric} that is better suited than the previous Podolsk\'y-Vr\'atn\'y parametrization \cite{PodolskyVratny:2023}, which we employed in the paper \cite{FernandezPodolskySenovilla:2024}.

\section{The case ${\Lambda>0}$: The super-Poynting vector and energy}
\label{section:sP-for-Lambda>0}

To analyze the character of gravitational radiation in the family of black hole spacetimes \eqref{ds2_simpl} with a positive cosmological constant ${\Lambda>0}$, it is important to evaluate the asymptotic super-Poynting vector field
\begin{equation}
	 \mathbf{\overline{P}} = -9\sqrt{2}\,|\phi_2|^2\,(1+2c\bar{c})
     \Big[ (1+c\bar{c})(c\, \mathbf{m} + \bar{c}\, \mathbf{\bar{m}})
     +(1+c\bar{c})\,c\bar{c}\, b^2\, \mathbf{k} + \frac{c\bar{c}}{b^2}\, \mathbf{l}\, \Big]\,.
    \label{super-Poyntin-def}
\end{equation}
This formula is the explicit combination of the expressions (B11), (B3), (B4), (B6) of \cite{FernandezPodolskySenovilla:2024} valid for spacetimes of algebraic type D. Here ${(\mathbf{k},\mathbf{l},\mathbf{m})}$ is the null tetrad on $\scri$ that is future-oriented and aligned with the two double PNDs. We have derived it explicitly in Eq.~\eqref{nulltetrad-Q<0}.
The coefficients $c$ and $b$ in \eqref{super-Poyntin-def} are defined as the projections of these vectors to the future-pointing \emph{unit timelike normal to}~$\scri$, namely ${c:=\sqrt{2}\,\, \mathbf{\bar{m}} \cdot \mathbf{n}}$ and ${b^{-2}:=-\sqrt{2}\,\, \mathbf{k} \cdot \mathbf{n}}$, see expressions (B1) and (B2) in \cite{FernandezPodolskySenovilla:2024}. In view of \eqref{eq:unit-normal-scri} and \eqref{norm-N_alpha}, these definitions can be equivalently written as the contractions
\begin{equation}
 c      = \sqrt{\frac{3}{\Lambda}}\, \sqrt{2}\,\,\langle \dd\Omega , \mathbf{\bar{m}}   \rangle \,,\qquad
 b^{-2} =-\sqrt{\frac{3}{\Lambda}}\, \sqrt{2}\,\,\langle \dd\Omega , \mathbf{k}   \rangle\,.
    \label{c-and-b-def}
\end{equation}
Using  \eqref{d Omega}, \eqref{nulltetrad-Q<0}, and \eqref{TRdef} it is now trivial to evaluate that\footnote{Notice that the key parameter $c$ vanishes if and only if the acceleration $\alpha$ is zero.}
\begin{equation}
 c      = - \alpha\,\,\im\,\sqrt{\frac{3}{\Lambda}}\, \frac{\sqrt{{P}}}{\rho} \,,\qquad
 b^{-2} = \sqrt{\frac{3}{\Lambda}}\, \frac{\sqrt{-{Q}}}{\rho} \,.
    \label{c-and-b}
\end{equation}
Employing the important relation \eqref{Q-on-scri-short} on scri we obtain
\begin{equation}
 c\bar{c}     = \alpha^2\,\frac{3}{\Lambda}\,\frac{{P_\scrisub}}{\rho^2_\scrisub} \,,\qquad
 b^2 = \sqrt{1+\alpha^2\,\frac{{P_\scrisub}}{{Q_\scrisub}}}\,,
    \label{c-and-b-other}
\end{equation}
and
\begin{equation}
 1+ c\bar{c}  = \frac{3}{\Lambda}\,\frac{(-{Q_\scrisub})}{\rho^2_\scrisub} >0 \,,\qquad
 1+ 2c\bar{c} = \frac{3}{\Lambda}\,\frac{\alpha^2{P_\scrisub}-{Q_\scrisub}}{\rho^2_\scrisub} >0\,.
    \label{c-consequently}
\end{equation}
Substituting these coefficients and the vectors \eqref{nulltetrad-Q<0} into \eqref{super-Poyntin-def} we get
\begin{equation}
 \mathbf{\overline{P}} = 18 \Big(\frac{3}{\Lambda}\Big)^{\frac{5}{2}}\,\,\alpha\,\,({Q}-\alpha^2{P})_\scrisub\,
    \Big(\frac{{P}\,{Q}}{\rho^6 }\Big)_\scrisub\,|\phi_2|^2\,
     \big(\,\partial_x + \alpha\,\partial_q\,\big).
    \label{super-Poynting}
\end{equation}

The pull-back on scri $\scri$ (see \cite{FernandezPodolskySenovilla:2024} for the details of this procedure) gives us the compact and simple formula for the asymptotic super-Poynting vector:
\begin{equation}
 \mathbf{\overline{P}} = \alpha\,|\phi_2|^2\,R\,\,\partial_{\bar x}\,,
    \label{super-Poynting-scri}
\end{equation}
where
\begin{align}
R(\bar x) := 18 \Big(\frac{3}{\Lambda}\Big)^{\frac{5}{2}}
    \,\big({Q}_{\!\scrisub}-\alpha^2 {P}_{\!\scrisub} \big)\,
    \frac{{P}_{\!\scrisub} {Q}_{\!\scrisub}}{\rho^6_\scrisub }\, . \label{def-R}
\end{align}
Recall that the functions $\rho_{\!\scrisub}$, ${P}_{\!\scrisub}$, ${Q}_{\!\scrisub}$ are given by  \eqref{rho2_on_scri}, \eqref{P_on_scri}, \eqref{Q_on_scri}, respectively, and $\bar x$ is the coordinate on $\scri$ related to $x$. Notice also that ${\mathbf{r} \propto \partial_x}$, see \eqref{Q<0-orthonormal}, and thus ${\mathbf{\overline{P}} \propto \mathbf{r}}$. It means that on $\scri$ the asymptotic super-Poynting vector is oriented (only) along the spatial direction determined by $\mathbf{r}$.

More importantly, it can be explicitly seen from \eqref{super-Poynting-scri} that ${\mathbf{\overline{P}} = 0 \Leftrightarrow  \alpha =0 }$.\footnote{Unless ${\phi_2=0}$ or ${R=0}$, but these conditions cannot be satisfied for black holes (they have ${m\ne0}$ and thus ${\phi_2\ne0}$) away from the axis of symmetry (where ${P\ne0}$, and also ${Q<0}$).} It means, that the \emph{black holes emit gravitational radiation if, and only if, they accelerate}.

It can also be observed that the timelike unit vector ${\mathbf{t} \propto \partial_q}$ given by  \eqref{Q<0-orthonormal} is not collinear with the unit vector~$\mathbf{n}$ \eqref{eq:unit-normal-scri-explicit} normal to~$\scri$, unless ${\alpha=0}$. In fact, for ${\alpha\ne0}$ the normal~${{\bf n}}$ is not coplanar with the two PNDs. As observed by Fern\'andez-\'Alvarez and Senovilla, following their Remark IV.4 in \cite{FernandezSenovilla:2022b}, this purely geometrical fact already implies that there is gravitational radiation arriving at $\scri$ only if the black holes are accelerating (see the related work \cite{FernandezSenovilla:2026b} for more details).

Using \eqref{c-and-b-other} and \eqref{c-consequently} we can also easily calculate the asymptotic super-energy density given by equation (B.13) in \cite{FernandezPodolskySenovilla:2024} as ${{\cal W}=6|\phi_2|^2 \big(1+6c\bar{c}(1+c\bar{c})\big)}$. It reads
\begin{equation}
{\cal W}=6|\phi_2|^2 \,\Big[1-6\,\alpha^2\Big(\frac{3}{\Lambda}\Big)^2\,
    \frac{{P}_{\!\scrisub} {Q}_{\!\scrisub}}{\rho^4_\scrisub } \Big].
\label{super-energy-scri}
\end{equation}
For vanishing acceleration~$\alpha$, this is just the constant ${{\cal W}=6|\phi_2|^2} $.

To explicitly complete the derivation, it remains to calculate the coefficient $\phi_2$. This can be obtained by rescaling the Weyl scalar $\Psi_2$ of the physical metric using the relation ${\phi_2=\Psi_2\,\Omega^{-3}}$, where ${\Omega=q-\alpha\,x}$, and evaluating it on $\scri$ (or directly from the unphysical metric and frames). It depends only on the physical parameters of the black holes, namely $m, \alpha, a, l, e, g, \Lambda$. In full generality, it is quite complicated but there are considerable simplifications in special subcases.

In the important case of vanishing acceleration (${\alpha=0}$) we get
\begin{eqnarray}
 \Psi_2 = \frac{-m + \im\,l\, (1-\tfrac{1}{3}\Lambda l^2)}{[1-\im\,q\,(l+a\,x)]^3}\,q^3
    +\frac{e^2+g^2}{[1 + q^2(l+a\,x)^2][1-\im\,q\,(l+a\,x)]^2}\,q^4 \,,  \label{Psi2-alpha=0}
\end{eqnarray}
and the conformal factor \eqref{Omega_new} is simply ${\Omega=q}$, so that ${\phi_2=\Psi_2/q^3}$. The conformal infinity $\scri$ is located at ${q=0}$, so that the second term proportional to ${e^2+g^2}$ vanishes there, and the first term reduces to
\begin{equation}
\phi_{2} = -m + \im\,l\, (1-\tfrac{1}{3}\Lambda l^2)\,.
\end{equation}
For ${l=0}$, that is for any Kerr-Newman-(anti-)de~Sitter black hole, we get simply ${\phi_{2} = -m }$, while for any Kerr-Newman-NUT black hole we get ${\phi_{2} = -m + \im\,l}$, so that ${{\cal W}=6(m^2+l^2)}$.

Complementarily, for all black holes without the NUT parameter (${l=0}$) we obtain
\begin{eqnarray}
 \Psi_2 = \bigg[-m \,\frac{1+\im\,\alpha a}{(1-\im\,q\,a\,x)^3}
    +\frac{(e^2+g^2)(q+\alpha\,x)}{(1 + q^2a^2x^2)(1-\im\,q\,a\,x)^2}\bigg] (q-\alpha\,x)^3 \,.  \label{Psi2-l=0}
\end{eqnarray}
and thus
\begin{equation}
\phi_{2} = -m \,\frac{1+\im\,\alpha a}{(1-\im\,\alpha a\, x^2 )^3}
    +\frac{(e^2+g^2)\,2\alpha \,x}{(1 + \alpha^2a^2\,x^4)(1 - \im\,\alpha a\, x^2 )^2}\,,
\end{equation}
which reduces to
\begin{equation}
\phi_{2}= -m \,\frac{1+\im\,\alpha a}{(1-\im\,\alpha a\, x^2 )^3}\,,
\end{equation}
when ${e=0=g}$, that is for accelerating uncharged Kerr-(anti-)de~Sitter black holes. For the classic C-metric without rotation (${a=0}$), it is just ${\phi_{2}=-m}$, implying ${{\cal W}=6m^2}$.

\section{The case ${\Lambda<0}$: The super-Poynting vector and energy}
\label{section:sP-for-Lambda<0}

In this final section we will investigate the gravitational radiation in the more complicated ${\Lambda<0}$ case of black holes in anti-de~Sitter universe. We have to distinguish two types of the regions on~$\scri$.

\subsection{Stationary regions ${Q_\scrisub >0}$}
\label{section:sP-for-Lambda<0-and-Q>0}

In the stationary regions ${Q_\scrisub >0}$ of the ${\Lambda<0}$ spacetime close to the anti-de~Sitter-like~$\scri$, we can apply the  null tetrad adapted to the two PNDs given by \eqref{nullframe-Q>0}. Moreover, we derived in \eqref{eq:unit-normal-Q>0} that the spacelike normal~$\mathbf{n}$, expressed in such a tetrad, reads
\begin{equation}	\label{eq:unit-normal-Q>0-again}
 \mathbf{n} =\sqrt{-\frac{3}{\Lambda}}\,\,\frac{1}{\rho_\scrisub}\,
     \,\Big( -\sqrt{{Q}_\scrisub}\, \mathbf{q} + \alpha\,\sqrt{{P}_\scrisub}\,\mathbf{r} \Big) \,.
\end{equation}
The (outwards oriented) spatial unit vector $\mathbf{q}$ determined by the plane of the two PNDs $\mathbf{k}$ and $\mathbf{l}$  \eqref{nullframe-Q>0}, which are spanned only by $\textbf{q}$
and $\textbf{T}$, is \emph{not} normal to $\scri$ because \eqref{eq:unit-normal-Q>0-again} also contains~$\mathbf{r}$  (unless ${\alpha=0}$). According to Remark IV.4 in \cite{FernandezSenovilla:2022b} it means that such black holes generate gravitational radiation if and only if they are accelerating.

Let us make this statement precise by explicitly calculating the corresponding super-Poynting vector $\mathbf{\overline{P}}$. We start by performing a simple rotation in the spatial 2-space spanned by the Cartesian vectors $\mathbf{q}$ and $\mathbf{r}$ of \eqref{Q>0-orthonormal}, namely
\begin{eqnarray}
	\tilde{\mathbf{q}} \rovno \cos\beta\,\textbf{q} - \sin\beta\,\textbf{r}\,, \nonumber \\
	\tilde{\mathbf{r}} \rovno \sin\beta\,\textbf{q} + \cos\beta\,\textbf{r}\,,  \label{q-r-rotation-Q>0}
\end{eqnarray}
where the angular parameter $\beta$ is defined as
\begin{equation}
	\sin\beta \defeq  \alpha\,\sqrt{-\frac{3}{\Lambda}}\,\,\frac{\sqrt{{P}_\scrisub}}{\rho_\scrisub}
     \qquad \Rightarrow \qquad
\cos\beta =  \sqrt{-\frac{3}{\Lambda}}\,\,\frac{\sqrt{{Q}_\scrisub}}{\rho_\scrisub}\,.  \label{Q>0-betal-def}
\end{equation}
It follows that on $\scri$ there is
\begin{equation}
	\tilde{\mathbf{q}}=-{\mathbf{n}}\,,\qquad
    \tilde{\mathbf{r}}\cdot\tilde{\mathbf{r}}=1\,,\qquad
    \tilde{\mathbf{r}}\cdot\mathbf{n}=0\,,  \label{Q>0:q=-n}
\end{equation}
so that $\tilde{\mathbf{q}}$ is the outwards oriented spatial unit vector \emph{normal} to~$\scri$. On the other hand, $\tilde{\mathbf{r}}$ is the spatial unit vector \emph{tangent} to~$\scri$. Together with the normalized and mutually orthogonal vectors $\mathbf{s}$ and $\mathbf{t}$, also tangent to~$\scri$, they form the basis of all vectors on scri.

In particular, \emph{any timelike unit vector~$\mathbf{u}$} on this Lorentzian (1+2) $\scri$ can be naturally expressed as
\begin{equation}	\label{eq:general-unit-timelike-Q>0}
 \mathbf{u} = \cosh\psi\,\mathbf{t} + \sinh\psi\,(\cos\phi\,\mathbf{s}-\sin\phi\,\tilde{\mathbf{r}})\,,
\end{equation}
because ${\mathbf{u}\cdot\mathbf{u}=-1}$ for \emph{all} values of the \emph{rapidity} (boost) parameter ~${\psi\in[0,\infty)}$ with respect to $\mathbf{t}$, and all values of the \emph{rotation}~${\phi\in[0,2\pi)}$. Actually, such a parametrization was conveniently employed in our previous works \cite{PodolskyOrtaggioKrtous:2003,KrtousPodolsky:2004,KrtousPodolsky:2005,PodolskyKadlecova:2009} investigating the relation between the type~D structure of principal null directions and the anti-de~Sitter-like scri, and the corresponding gravitational radiation.

This explicit expression of $\mathbf{u}$ can now be compared to the general formula (B.1) given in Appendix~B of \cite{FernandezSenovilla:2026}, namely
\begin{equation}
\mathbf{u} = a\,\mathbf{k}  +b\,\mathbf{l} + d\,\mathbf{m} + \bar{d}\,\mathbf{\bar{m}}\,.
\label{general-oberver-u}
\end{equation}
Using the standard relations
\begin{equation}
	\mathbf{k} =  \tfrac{1}{\sqrt2} (\mathbf{t} + \mathbf{q})\,, \qquad
	\mathbf{l} =  \tfrac{1}{\sqrt2} (\mathbf{t} - \mathbf{q})\,, \qquad
	\mathbf{m} =  \tfrac{1}{\sqrt2} (\mathbf{s} + \mathrm{i}\, \mathbf{r})\,,  \label{Q>0-orthonormal-def}
\end{equation}
and \eqref{q-r-rotation-Q>0}, by comparing \eqref{general-oberver-u} with \eqref{eq:general-unit-timelike-Q>0} we easily derive that
\begin{eqnarray}
	a \rovno \tfrac{1}{\sqrt{2}}\,(\cosh\psi - \sinh\psi\sin\phi\sin\beta) \,, \nonumber \\
	b \rovno \tfrac{1}{\sqrt{2}}\,(\cosh\psi + \sinh\psi\sin\phi\sin\beta)\,,  \label{abc-Q>0}\\
	d \rovno \tfrac{1}{\sqrt{2}}\,\sinh\psi\,(\cos\phi + \im\,\sin\phi\cos\beta) \,.\nonumber
\end{eqnarray}
It automatically satisfies the normalization constraint (B.2) of \cite{FernandezSenovilla:2026}, that is ${ab-d\bar{d} = \tfrac{1}{2}}$, and by construction it also satisfies the orthogonality condition ${\mathbf{u}\cdot\mathbf{n}=0}$, which is (B.3). Therefore, we can  directly employ the formula (B.58) valid for the type~D spacetimes, namely
\begin{equation}
\mathbf{\overline{P}}\cdot \mathbf{n}\,(\mathbf{u}) = 36 \,|\phi_2|^2\,(ab+d\bar{d}\,)\big( a\,B_1 + b\,B_2 \big),
\label{B.58}
\end{equation}
where now
\begin{equation}
B_1 := \mathbf{k}\cdot \mathbf{n}= -\tfrac{1}{\sqrt2} \,\cos\beta\,, \qquad
B_2 := \mathbf{l}\cdot \mathbf{n} = \tfrac{1}{\sqrt2} \,\cos\beta\,.
\end{equation}
We thus get
\begin{eqnarray}
\mathbf{\overline{P}}\cdot \mathbf{n}\,(\mathbf{u}) \rovno 18 \,|\phi_2|^2\,\sin\beta \cos\beta\nonumber\\
&& \times \big[ 1 + 2\sinh^2\psi\,(1-\sin^2\phi\sin^2\beta) \big]\sinh\psi\sin\phi\,.
\label{s-Poyinting-Q>0}
\end{eqnarray}
The first line determines the (positive) \emph{magnitude} of the super-Poynting vector $\mathbf{\overline{P}}$ projected onto the normal $\mathbf{n}\,(\mathbf{u})$ to scri. In view of the relations \eqref{Q>0-betal-def} we immediately obtain
\begin{eqnarray}
	18 \,|\phi_2|^2\,\sin\beta \cos\beta \rovno \alpha\,\bigg(\!\!-\frac{3}{\Lambda}\bigg)18 \,|\phi_2|^2\,\frac{\sqrt{{P}_\scrisub {Q}_\scrisub}}{\rho_\scrisub^2} \,,  \label{Q>0-Poynting-magnitude}
\end{eqnarray}
so it depends on the position of the point~${x\in\scri}$, but it is \emph{independent} of the choice of~$\mathbf{u}$. Notice that it vanishes on poles given by ${P_{\!\scrisub}=0}$,  and also on the ``horizon boundaries'' given by  ${Q_{\!\scrisub}=0}$.

The complete form of the super-Poynting vector projected onto the normal to~$\scri$ is
\begin{eqnarray}
\mathbf{\overline{P}}\cdot \mathbf{n}\,(\mathbf{u}) \rovno
18\,|\phi_2|^2\,\,\alpha\,\,\frac{\sqrt{P_\scrisub Q_\scrisub}}{Q_\scrisub\! + \alpha^2P_\scrisub}
\big[ 1 + 2\sinh^2\psi\,(1-\sin^2\phi\sin^2\beta) \big]\sinh\psi\sin\phi\,.
  \label{Q>0-Poynting-final}
\end{eqnarray}
It is now obvious that there is no radiation if and only if~${\alpha=0}$, that is \emph{if and only if the black holes do not accelerate}. In such a case ${\mathbf{\overline{P}}\cdot \mathbf{n}\,(\mathbf{u}) = 0}$ \emph{everywhere} in the region ${Q_\scrisub >0}$  for \emph{any} timelike observer~$\mathbf{u}$ (specified by $\psi,\phi$) on~the anti-de~Sitter-like~$\scri$.

The second line of \eqref{s-Poyinting-Q>0} expresses the dependence of~${\,\mathbf{\overline{P}}\cdot \mathbf{n}\,(\mathbf{u})}$ on the choice of the observer~$\mathbf{u}$ at~${x\in\scri}$. As given by \eqref{eq:general-unit-timelike-Q>0}, it is determined by the boost parameter ~${\psi\in[0,\infty)}$ and the rotational parameter~${\phi\in[0,2\pi)}$. Since the expression in the square brackets is always positive, we conclude that ${\mathbf{\overline{P}}\cdot \mathbf{n} > 0}$ for ${\phi\in(0,\pi)}$, while ${\mathbf{\overline{P}}\cdot \mathbf{n} > 0}$ for ${\phi\in(\pi,2\pi)}$. In the former case, it is viewed as an ingoing radiation, while in the latter case as outgoing.

It can be also seen from  \eqref{Q>0-Poynting-final} that ${\mathbf{\overline{P}}\cdot \mathbf{n} = 0}$ (there is no radiation flux) for ${\phi=0}$ and ${\phi=\pi}$, that is for \emph{special observers} such that ${\mathbf{u} = \cosh\psi\,\mathbf{t} + \sinh\psi\,\mathbf{s}}$, and ${\mathbf{u} = \cosh\psi\,\mathbf{t} - \sinh\psi\,\mathbf{s}}$, respectively. These have no components in the spatial direction~$\tilde{\mathbf{r}}$. And since (by our definition) they also have no components in the direction~${\tilde{\mathbf{q}}=-{\mathbf{n}}}$ normal to scri, they are \emph{linear combinations of the vectors} $\textbf{T}$ and $\textbf{R}$ which at any ${x=\hbox{const.}}$ are linear combinations of the two Killing vectors ${\partial_{t}}$ and ${\partial_{\varphi}}$, see \eqref{TRdef} and \eqref{Q>0-orthonormal} (notice however that in the general case the coefficients $A_\scrisub, B_\scrisub, C_\scrisub, D_\scrisub$ are functions of~$x$. In particular, this is the case of the privileged timelike observer ${\mathbf{u} = \mathbf{t} = \textbf{T}/(\sqrt{Q_\scrisub}\,\rho_\scrisub)
= ( A_\scrisub \, \partial_{t} - \alpha^2 B_\scrisub \, \partial_{\varphi})/(\sqrt{{Q}_\scrisub}\,\rho_\scrisub)}$  corresponding to ${\psi=0}$. For vanishing acceleration (${\alpha=0}$) it simplifies to ${\mathbf{u} = \sqrt{- \frac{3}{\Lambda}}\,\,\partial_{t}}$.

Actually, this argumentation can be reverted. \emph{Whenever} the observer $\mathbf{u}$ on scri has the component in the spatial direction~$\tilde{\mathbf{r}}$ that is \emph{pointing away from the 2-plane spanned by the two vectors}~${\mathbf{t} \propto \textbf{T}}$ and ${\mathbf{s} \propto \textbf{R}}$, it follows from \eqref{eq:general-unit-timelike-Q>0} that
${\,\sinh\psi\sin\phi \ne 0\,}$, and thus according to the explicit formula \eqref{Q>0-Poynting-final} we get ${\mathbf{\overline{P}}\cdot \mathbf{n}\,(\mathbf{u}) \ne 0}$ (unless ${\alpha=0}$). We conclude that \emph{general} observers on (and close to)~$\scri$ of accelerating black holes would detect gravitational radiation despite the fact that the region there is stationary (${Q_\scrisub >0}$). Only those observers aligned with the plane of the two vectors $\textbf{T}$ and $\textbf{R}$ would observe no radiation. The same is true for observers on the axes of symmetry (${P_{\!\scrisub}=0}$) and on the horizons (${Q_{\!\scrisub}=0}$). These conclusions are consistent because both the poles and the horizons are generated by the Killing vectors of the spacetime.

\subsection{Non-stationary regions ${Q_\scrisub <0}$}
\label{section:sP-for-Lambda<0-and-Q<0}

To explicitly express the super-Poynting vector using the formula \eqref{B.58},
we must find the general normalized timelike unit vector $\mathbf{u}$, as in \eqref{general-oberver-u}. To this end, we perform a suitable rotation of the orthonormal frame \eqref{Q<0-orthonormal2} analogous to \eqref{q-r-rotation-Q>0}. However, instead of performing a spatial rotation, now we need to do a suitable boost, namely
\begin{eqnarray}
	\tilde{\mathbf{t}} \rovno \cosh\gamma\,\textbf{t} + \sinh\gamma\,\textbf{r}\,, \nonumber \\
	\tilde{\mathbf{r}} \rovno \sinh\gamma\,\textbf{t} + \cosh\gamma\,\textbf{r}\,,  \label{q-r-rotation-Q<0}
\end{eqnarray}
where the parameter $\gamma$ is defined as
\begin{equation}
\sinh\gamma \defeq \sqrt{-\frac{3}{\Lambda}}\,\,\frac{\sqrt{-{Q}_\scrisub}}{\rho_\scrisub}
\qquad \Rightarrow \qquad
	\cosh\gamma = \alpha\,\sqrt{-\frac{3}{\Lambda}}\,\,\frac{\sqrt{{P}_\scrisub}}{\rho_\scrisub}
\,.  \label{Q<0-gamma-def}
\end{equation}
It follows that on $\scri$
\begin{equation}
	\tilde{\mathbf{r}}={\mathbf{n}}\,,\qquad
    \tilde{\mathbf{t}}\cdot\tilde{\mathbf{t}}=-1\,,\qquad
    \tilde{\mathbf{t}}\cdot\mathbf{n}=0\,,  \label{Q<0:r=n}
\end{equation}
so that $\tilde{\mathbf{r}}$ is the (inwards oriented) spatial unit vector \emph{normal} to~$\scri$, whereas $\tilde{\mathbf{t}}$ is the timelike unit vector \emph{tangent} to~$\scri$. Together with the normalized and mutually orthogonal vectors $\mathbf{s}$ and $\mathbf{q}$, also tangent to~$\scri$, they form the basis of all vectors on this ${Q_\scrisub <0}$ part of anti-de~Sitter-like scri. It means that \emph{any} timelike unit vector~$\mathbf{u}$ on~such Lorentzian $\scri$ can be expressed as
\begin{equation}	\label{eq:general-unit-timelike-Q<0}
 \mathbf{u} = \cosh\psi\,\tilde{\mathbf{t}} + \sinh\psi\,(\cos\phi\,\mathbf{s}-\sin\phi\,\mathbf{q})\,,
\end{equation}
because, as in the previous case \eqref{eq:general-unit-timelike-Q>0}, ${\mathbf{u}\cdot\mathbf{u}=-1}$ for all values of the rapidity parameter ~${\psi\in[0,\infty)}$ with respect to~${\tilde{\mathbf{t}}}$, and all values of the rotation~${\phi\in[0,2\pi)}$. By comparing \eqref{eq:general-unit-timelike-Q<0}
with \eqref{general-oberver-u}, that is ${\mathbf{u} = a\,\mathbf{k}  +b\,\mathbf{l} + d\,\mathbf{m} + \bar{d}\,\mathbf{\bar{m}}}$, we derive he expressions
\begin{eqnarray}
	a \rovno \tfrac{1}{\sqrt{2}}\,(\cosh\psi\cosh\gamma - \sinh\psi\sin\phi) \,, \nonumber \\
	b \rovno \tfrac{1}{\sqrt{2}}\,(\cosh\psi\cosh\gamma + \sinh\psi\sin\phi)\,,  \label{abc-Q<0}\\
	d \rovno \tfrac{1}{\sqrt{2}}\,(\sinh\psi\cos\phi  - \im\,\cosh\psi\sinh\gamma) \,.\nonumber
\end{eqnarray}
They satisfy the normalization condition ${ab-d\bar{d} = \tfrac{1}{2}}$ and also the orthogonality condition ${\mathbf{u}\cdot\mathbf{n}=0}$, i.e. the constraints (B.2) and (B.3) in \cite{FernandezSenovilla:2026}. We can thus again directly use the formula (B.58) therein, that is \eqref{B.58}, with
\begin{equation}
B_1 = -\tfrac{1}{\sqrt2} \,\sinh\gamma = B_2  \,.
\end{equation}
Straightforward evaluation gives us the explicit asymptotic super-Poynting vector
\begin{eqnarray}
\mathbf{\overline{P}}\cdot \mathbf{n}\,(\mathbf{u}) \rovno -18 \,|\phi_2|^2\,\sinh\gamma \cosh\gamma  \nonumber\\
&& \times \big[ \cosh^2\psi \cosh(2\gamma)+\sinh^2\psi\cos(2\phi) \big]\cosh\psi\,,
\label{s-Poyinting-Q<0}
\end{eqnarray}
where
\begin{eqnarray}
	18 \,|\phi_2|^2\,\sinh\gamma \cosh\gamma \rovno \alpha\,\bigg(\!\!-\frac{3}{\Lambda}\bigg)18 \,|\phi_2|^2\,\frac{\sqrt{-{P}_\scrisub {Q}_\scrisub}}{\rho_\scrisub^2}  \,. \label{Q<0-Poynting-magnitude2}
\end{eqnarray}
Notice that the second line of \eqref{s-Poyinting-Q<0} is always non-negative. We thus arrive at the result
\begin{eqnarray}
\mathbf{\overline{P}}\cdot \mathbf{n}\,(\mathbf{u}) \rovno
-18\,|\phi_2|^2\,\,\alpha\,\,\frac{\sqrt{-P_\scrisub Q_\scrisub}}{Q_\scrisub\! + \alpha^2P_\scrisub}
\big[ \cosh^2\psi \cosh(2\gamma)+\sinh^2\psi\cos(2\phi) \big]\cosh\psi\,,
  \label{Q<0-Poynting-final}
\end{eqnarray}
which is the analogue of \eqref{Q>0-Poynting-final}. It means that also in the non-stationary regions ${Q_\scrisub <0}$ the radiation vanishes if and only of ${\alpha=0}$, i.e. when the black holes do not accelerate.

\section{Conclusions}

We derived the new simple metric \eqref{A-new-metric}--\eqref{Q_new}, based on the recent convenient A$^+$ representation of all Pleba\'nski-Demia\'nski black holes with the Kerr rotation, NUT twist, charges, acceleration, an any value of the cosmological constant. We showed that it can be used to study the conformal infinity of the whole class, and thus can be useful for investigation of the asymptotic structure in the non-trivial cases ${\Lambda>0}$ and ${\Lambda<0}$. As a particular application, following the covariant and gauge-independent methods recently developed by Fern\'andez-\'Alvarez and Senovilla, we evaluated by analytical means the asymptotic super-Poynting vector and the superenergy density, see Eqs.~\eqref{super-Poynting-scri} and~\eqref{super-energy-scri} in the ${\Lambda>0}$ case, and Eqs.~\eqref{Q>0-Poynting-final} and~\eqref{Q<0-Poynting-final} in the ${\Lambda<0}$ case. This explicitly confirmed that the Pleba\'nski-Demia\'nski black holes emit gravitational radiation in spacetimes with any cosmological constant~$\Lambda$ if and only if they accelerate. And vice versa: the Fern\'andez-\'Alvarez-Senovilla criterion for the presence of gravitational radiation justifies the physical interpretation of the parameter $\alpha$ in the A$^+$ form of the Pleba\'nski-Demia\'nski spacetimes as representing acceleration of the black hole.

Ii is our hope that the metrics \eqref{A-new-metric} and \eqref{conformal-metric-final}, with the related canonical frames on the conformal infinity presented in Sec.~\ref{section:PND-Weyl-for-Lambda>0} and Sec.~\ref{section:PND-Weyl-for-Lambda<0}, will find other applications in more detailed studies of the asymptotic structure of this famous class of black holes, including the specific gravitational radiation they emit.

\section*{Acknowledgments}

This work was supported by the Czech Science Foundation Grant No.~GA\v{C}R 26-22381S. I also thank Francisco Fern\'andez-\'Alvarez and Jos\'e~M.~M.~Senovilla for many discussions and useful feedbacks.

\section*{Data availability}

No data were created or analyzed in this study

\end{document}